\documentclass[
reprint,
preprintnumbers,
nofootinbib,
 amsmath,amssymb,
 aps,
]{revtex4-2}

\usepackage{graphicx}
\usepackage{dcolumn}
\usepackage{bm}
\usepackage{url}
\usepackage{hyperref}
\usepackage{amscd, mathrsfs}
\usepackage{amsthm}
\usepackage{cases}
\usepackage{here}
\usepackage{bm}
\usepackage{graphics}
\usepackage{ulem}
\usepackage{multirow}
\usepackage{subcaption}

\usepackage{color}
\definecolor{moss}{cmyk}{0.41, 0.27, 0.45, 0.4}
\hypersetup{
 colorlinks=true,
 urlcolor=moss,
 pagecolor=moss,
 menucolor=moss,
 linkcolor=moss,
 filecolor=moss,
 citecolor=moss
}

\begin{document}

\preprint{RESCEU-9/22, Prepared for submission to Physics Letters B}

\title{New description of the scaling evolution of the\\cosmological magneto-hydrodynamic system}

\author{Fumio Uchida}\email{fuchida@resceu.s.u-tokyo.ac.jp}
\author{Kohei Kamada}

\author{Jun'ichi Yokoyama}
\altaffiliation[Also at]{
    Kavli Institute for the Physics and Mathematics of the Universe (Kavli IPMU),\\WPI, UTIAS, The University of Tokyo, Kashiwa, Chiba 277-8568, Japan,\\
    Trans-scale Quantum Science Institute,\\The University of Tokyo, Tokyo 113-0033, Japan}
\affiliation{
    Research Center for the Early Universe (RESCEU) and Department of Physics,\\ Graduate School of Science, The University of Tokyo, Tokyo 113-0033, Japan
}
\author{Motoko Fujiwara}
\affiliation{
    Physik-Department, 
    Technische Universit\"{a}t M\"{u}nchen,
    James-Franck-Str.~1 85748 Garching, Germany
}
\date{\today}

\begin{abstract}
    We present a new description of cosmological evolution of the primordial magnetic field under the condition that it is non-helical and its energy density is larger than the kinetic energy density.
    We argue that the evolution can be described by  four different regimes,
    according to whether the decay dynamics is linear or not, and whether the dominant dissipation term is the shear viscosity or the drag force.
    Using this classification and conservation of the Hosking integral, we present analytic models to adequately interpret the results of various numerical simulations of field evolution with variety of initial conditions.
    It is found that, contrary to the conventional wisdom, the decay of the field is generally slow, exhibiting the inverse transfer, because of the conservation of the Hosking integral.
    Using the description proposed here, we may trace the intermediate evolution history of the magnetic field and clarify whether each process governing its evolution is frozen or not, which is essential to follow the evolution of relatively weak magnetic fields.
\end{abstract}

\maketitle

\section{\label{sec:Introduction}Introduction}
The cosmological magnetic field can serve as a probe into the early universe.
Observations tell that the intergalactic magnetic field exists even in the void \cite{NeronovVovk10, 2010MNRAS.406L..70T, 2010ApJ...722L..39A, 2011MNRAS.414.3566T, 2011ApJ...727L...4D, 2011APh....35..135E, 2011A&A...529A.144T, 2011ApJ...733L..21D, 2012ApJ...747L..14V, Takahashi:2013lba, 2015ApJ...814...20F, Ackermann+18, 2020ApJ...902L..11A}, and its origin may be attributed to the early universe.
Possible formation scenarios include magnetogenesis during or at the end of inflation \cite{1988PhRvD..37.2743T, 1992ApJ...391L...1R, 1992PhRvD..46.5346G, 1995PhRvD..52.6694M, PhysRevD.77.123002, 2008JCAP...04..024B, 1995PhRvD..52.1955L, 1995PhRvL..75.3796G, 2004PhRvD..69d3507B, 2007JCAP...02..030B, martin2008generation, 2009JCAP...08..025D, 2009JCAP...12..009K, 2014JCAP...05..040K, 1998PhRvD..57.7139C, 2000PhRvD..62j3512G, 2001PhLB..501..165D, 2002PhRvD..65f3505D, 1993PhRvD..48.2499D, 1999PhLB..455...96B, 2005PhRvD..71j3509A, 2008PhRvL.100x1301D, 2006JCAP...10..018A, 2011JCAP...03..037D, 2015JCAP...05..054F, 2016JCAP...10..039A} and at the phase transitions \cite{PhysRevLett.51.1488, 1989ApJ...344L..49Q, PhysRevD.50.2421, 1991PhLB..265..258V, 1993PhLB..319..178E, PhysRevD.58.103505, PhysRevD.53.662, PhysRevD.54.1291, PhysRevD.55.4582,  PhysRevD.57.664, PhysRevD.56.6146, PhysRevLett.79.1193, PhysRevD.62.103008, PhysRevLett.87.251302}.
To understand the role of the primordial magnetic field
and to connect the models of the primordial universe and observation, 
describing its magneto-hydrodynamic (MHD) evolution within the framework of the standard cosmology is the first and the most important step.

In this context, Banerjee \& Jedamzik~\cite{Banerjee+04} tried to describe the evolution of the cosmological magnetic field comprehensively, which
has been commonly accepted. However, some of their results appear to be inconsistent with those of recent numerical MHD simulations \cite{2014ApJ...794L..26Z, 2015PhRvL.114g5001B, Brandenburg:2016odr, Brandenburg+17, 2017PhRvE..96e3105R, 2017MNRAS.472.1628P}.
As a representative scenario, let us assume that the coupled system of the primordial magnetic field and the plasma filling the universe is {\it magnetically dominant} and {\it non-helical} initially, {\it e.g.}, at the electroweak symmetry breaking temperature.
While the conventional analysis assumes that the magnetic energy of the long-wavelength modes is intact as the short-wavelength modes decay, numerical simulations exhibit the so-called {\it inverse transfer} \cite{2014ApJ...794L..26Z, 2015PhRvL.114g5001B, Brandenburg:2016odr, Brandenburg+17, 2017PhRvE..96e3105R, 2017MNRAS.472.1628P}, in which the long-wavelength modes are enhanced in contrast to the decaying short-wavelength modes.
In this letter, we present analytic description of the evolution of the cosmological magnetic field with the initial conditions stated above, consistently with the numerical results.

The system during its evolution can be classified into four different regimes, according to whether the dynamics is linear or nonlinear with respect to the magnetic field and whether the kinetic dissipation is mainly due to shear viscosity or drag force.
The evolution of cosmological magnetic field should inevitably experience these different regimes since the dominant source of dissipation in the early universe vary with time among the collision between the particles of the fluid, the friction from the background free-streaming particles such as photons and neutrinos, and the Hubble friction in the matter dominated era.
As for the linear regimes with either shear viscosity or drag force, our analysis is partly based on Ref.~\cite{Banerjee+04}, and for the nonlinear regime with shear viscosity, mostly on Ref.~\cite{Hosking+21, Hosking+22}, which resolves the inconsistency between theory and numerical studies by introducing the {\it Hosking integral} (or the {\it Saffman helicity invariant}) \cite{Hosking+21} and the {\it reconnection time scale}.
The turbulent regime with drag force has never been discussed in literature, on which an analysis is provided in this letter consistently with the other three regimes.
For the first time, we integrate the analyses in these four regimes, which complete the description of the evolution of the magnetically-dominant non-helical cosmological magnetic field throughout the cosmic history from its generation to the present.
While the analysis in Ref.~\cite{Hosking+21, Hosking+22} may be sufficient to connect the properties of magnetic field in the early  universe and at present for relatively strong one,
our analysis is essential for its complete history.
To see the difference clearly, we focus on the case of weak magnetic field where the reconnection is driven by the Sweet--Parker reconnection~\cite{sweet58, Parker57}.

\section{\label{sec:setup}Equations of magneto-hydrodynamics}
We consider a coupled system of the magnetic field and the velocity field of the plasma fluid in the early universe.
In the following expressions, $\tau$ is the conformal time, $\bm{x}$ is the comoving coordinate, and $\rho$ and $p$ are the comoving energy density and the comoving pressure of the fluid, respectively.
The comoving quantities are related to the physical fields denoted by a subscript $_\text{p}$ as
\begin{eqnarray}
    {\bm B} = a^2 {\bm B}_\text{p},\quad 
    {\bm v} = {\bm v}_\text{p},\quad
    \rho = a^4\rho_\text{p},\quad
    p = a^4 p_\text{p},\notag\\
    \sigma = a \sigma_\text{p},\quad
    \eta = a^{-1}\eta_\text{p},\quad
    \alpha = a \alpha_\text{p},
    \label{eq:Relations of comoving and physical quantities}
\end{eqnarray}
where $a$ is the scale factor of the universe.
The equations of motions are
\begin{eqnarray}
    \partial_\tau {\bm B} -{\bm \nabla} \times ({\bm v} \times {\bm B})= \frac{1}{\sigma} {\bm \nabla}^2 {\bm B}
    \label{eq:Faraday's induction equation}
\end{eqnarray}
for the magnetic field $\bm{B}(\tau,\bm{x})$ and
\begin{eqnarray}
    \partial_\tau {\bm v} +({\bm v} \cdot {\bm \nabla}){\bm v}
    -\frac{1}{\rho+p}({\nabla} \times {\bm B}) \times {\bm B}+\frac{1}{\rho+p}{\bm \nabla} p \notag\\
    =\eta \left[{\bm \nabla}^2 {\bm v} + \frac{1}{3} {\bm \nabla}({\bm \nabla} \cdot {\bm v})\right]-\alpha\bm{v}
    \label{eq:Navier--Stokes equation}
\end{eqnarray}
for the velocity field $\bm{v}(\tau,\bm{x})$.
The right hand sides of these equations \eqref{eq:Faraday's induction equation} and \eqref{eq:Navier--Stokes equation} represent the dissipation of the energy, in terms of the electric conductivity, $\sigma$, the shear viscosity, $\eta$, and the drag force coefficient, $\alpha$.
The former two quantities originate from the collision between constituent particles of the plasma, while the last one from the background of the system, {\it i.e.,} the Hubble friction and/or free-streaming particles.
As the temperature of the universe decreases, the dominant term for the dissipation changes.
These equations are closed together with the continuity equation
\begin{eqnarray}
    \partial_\tau \rho +{\bm \nabla} \cdot[(\rho+p){\bm v} ]=\bm{E}\cdot(\bm{\nabla}\times\bm{B})\left(+aH\rho\right)&,
    \label{eq:Continuity equation}\\
    \text{where}\quad\bm{E}=\left(\frac{1}{\sigma}\bm{\nabla}-\bm{v}\right)\times\bm{B}&.
    \label{eq:Electric field in terms of magnetic field}
\end{eqnarray}
The last term in the right-hand-side of Eq.~\eqref{eq:Continuity equation} is present in the matter dominated era.

We assume the homogeneity and isotropy on average and express the magnetic and velocity field configurations on each time slice by a few parameters, by treating them as stochastic fields.
If we look inside a sufficiently small region, the magnetic field will be almost coherent and it appears anisotropic.
However, the present Hubble patch is composed of many coherent subpatches that have random magnetic fields assigned independently by a single probability distribution.
The universality of the probability distribution throughout the space guarantees the homogeneity, and the isotropy is imposed on the probability distribution.
For later purpose, we further assume that the probability distribution is almost Gaussian.
On top of these assumptions, 
we parametrize the magnetic field by its typical strength, $B$, and coherence length, $\xi_{\text{M}}$.
One may define these quantities in terms of the power spectrum, $P_B(k,\tau)$,
of the magnetic field,
\begin{equation}
    \langle\boldsymbol{B}(\boldsymbol{k})\cdot\boldsymbol{B}(\boldsymbol{k}') \rangle=P_B(k,\tau)(2\pi)^3\delta^3(\boldsymbol{k}+\boldsymbol{k}'),
\end{equation}
where $\boldsymbol{B}(\boldsymbol{k})$ denotes a Fourier mode.
\begin{align}
    B^2&:=\langle \bm{B}^2\rangle=\int \frac{d^3k}{(2\pi)^3}P_B(k,\tau),\\
    \xi_{\text{M}}&:=\frac{1}{B^2}\int \frac{d^3k}{(2\pi)^3}\frac{2\pi}{k}P_B(k,\tau) ,
    \label{eq:Definition of the parameters}
\end{align}
In the same way, we can define the typical velocity, $v$, and the coherence length of the velocity field, $\xi_{\text{K}}$, in tems of the velocity power spectrum, $P_v(k,\tau)$, as
\begin{align}
    v^2&:=\langle \bm{v}^2\rangle=\int \frac{d^3k}{(2\pi)^3}P_v(k,\tau),\\
    \xi_{\text{M}}&:=\frac{1}{v^2}\int \frac{d^3k}{(2\pi)^3}\frac{2\pi}{k}P_v(k,\tau) .
    \label{eq:Definition of the parameters2}
\end{align}
In the rest of the letter, we are going to determine the time dependence of these quantities,
which shows scaling behavior. \\

Conserved quantities play important roles to determine the scaling evolution of the magnetic and velocity
fields.
In magnetically dominant regimes, where the energy density of the non-helical magnetic field is dominant over that of the velocity field,
an approximate conserved quantity called Hosking integral,
\begin{eqnarray}
    I_{H_{\text{M}}}
    :=\int_{V} d^3r \langle h_{\text{M}}(\bm{x})h_{\text{M}}(\bm{x}+\bm{r})\rangle,
    \label{eq:Definition of Saffmnan helicity integral}\\
    \text{where}\quad
    h_{\text{M}}:=\bm{A}\cdot\bm{B},
    \label{eq:Definition of helicity density}
\end{eqnarray}
has been recently proposed~\cite{Hosking+21}. 

Here integral is taken over a volume $V$ much larger than the correlation volume, $\xi_\text{M}^3$, of the magnetic field.  Since the correlation length of $h_{\text{M}}$ is also expected to be of order of  $\xi_\text{M}$, 
on dimentional grounds we find
\begin{eqnarray}
    I_{H_{\text{M}}}\simeq B^4 \xi_\text{M}^5
\end{eqnarray}
up to a spectrum-dependent numerical factor.
In particular, we obtain the following constraint between the parameters at the temperature of the universe, $T$.
\begin{eqnarray}
    B(T)^4 \xi_\text{M}(T)^5=B_\text{ini}^4 \xi_\text{M,ini}^5,
    \label{eq:Condition from conservation of the Hosking integral}
\end{eqnarray}
where the subscript $_\text{ini}$ denotes the initial condition.
Indeed, the recent numerical study confirms that the Hosking integral is well conserved~\cite{zhou2022scaling}.
This conservation law restricts the evolution of the system as long as the system is magnetically dominant.
We will determine the scaling exponents using Eq.~\eqref{eq:Condition from conservation of the Hosking integral} for each regime in the succeeding section.

\section{\label{sec:Regime-dependent analysis}Regime-dependent analyses}
We now turn to the regime-dependent analysis, which is the main result of this letter.
As a preparation, we define a quantity that determines the boundary between non-linear and linear regimes by comparing the contributions from non-linear and linear terms in the induction equation, \eqref{eq:Faraday's induction equation}.
From rough estimates, $\vert{\bm \nabla} \times ({\bm v} \times {\bm B})\vert \sim v B/\min\{\xi_\text{M}, \xi_\text{K}\}$ and $ \vert{\bm \nabla}^2 {\bm B}/\sigma \vert \sim B/\sigma \xi_\mathrm{M}^2$, we define the magnetic Reynolds number,
\begin{eqnarray}
    \text{Re}_{\text{M}}
    :=\frac{\sigma v \xi_\text{M}^2}{\min\{\xi_\text{M}, \xi_\text{K}\}}. 
    \label{eq:Definition of Magnetic Reynolds number}
\end{eqnarray}
We identify $\text{Re}_\text{M}>(<)1$ with the non-linear (linear) regimes.

We also define another quantity that determines the boundary between viscous and dragged regimes by comparing the contributions in the right-hand-side of the Navier--Stokes equation, Eq.~\eqref{eq:Navier--Stokes equation}, at the scale of the kinetic coherence length, $\xi_\text{K}$.
By estimating that $\alpha |{\bm v}| \sim \alpha v$ and 
$\left\vert\eta \left[{\bm \nabla}^2 {\bm v} + \frac{1}{3} {\bm \nabla}({\bm \nabla} \cdot {\bm v})\right]\right\vert \sim \eta v/\xi_\mathrm{K}^2$,
we define the quantity that characterises the ratio of dissipation terms as 
\begin{eqnarray}
    r_\text{diss}:=\frac{\alpha\xi_\text{K}^2}{\eta}. 
    \label{eq:Ratio of the dissipation terms}
\end{eqnarray}
and then $r_\text{diss}<(>)1$ corresponds to the viscous (drag) regimes.
In the following, we study the evolution of the system by classifying the regimes according to the criteria determined by these quantities.
One can find the results of our analysis for each regime in Table~\ref{tab:summary of the decay laws}.

\subsection{\label{sec:Nonlinear with shear viscosity}Nonlinear regime with shear viscosity}
First, we consider the case where the magnetic Reynolds number is larger than unity and the shear viscosity is dominant over the drag force, 
\begin{eqnarray}
    \text{Re}_\text{M}\gg1,\quad r_\text{diss}\ll1,
    \label{eq:Conditions of nonlinear regime with shear viscosity}
\end{eqnarray}
In this regime, the evolution of the system is determined by the quasi-stationary condition of the magnetic reconnection with dissipation due to the shear viscosity, known as the Sweet--Paker reconnection~\cite{sweet58, Parker57}, which drives the transfer between the magnetic and kinetic  energy, if the magnetic field is not so strong (or more precisely if the {\it Lundquist number} is not so large)\footnote{
For stronger magnetic field, we expect that the reconnection is driven by the fast reconnection \cite{ji2011phase}.
In this case, the analysis in Ref.~\cite{Hosking+21, Hosking+22} may be sufficient.}.
The application of this physical mechanism to this regime is originally discussed in Ref.~\cite{Hosking+21}. Based on their discussion, we derived the scaling laws for MHD system.

\begin{figure*}[thb]
\includegraphics[width=0.8\textwidth]{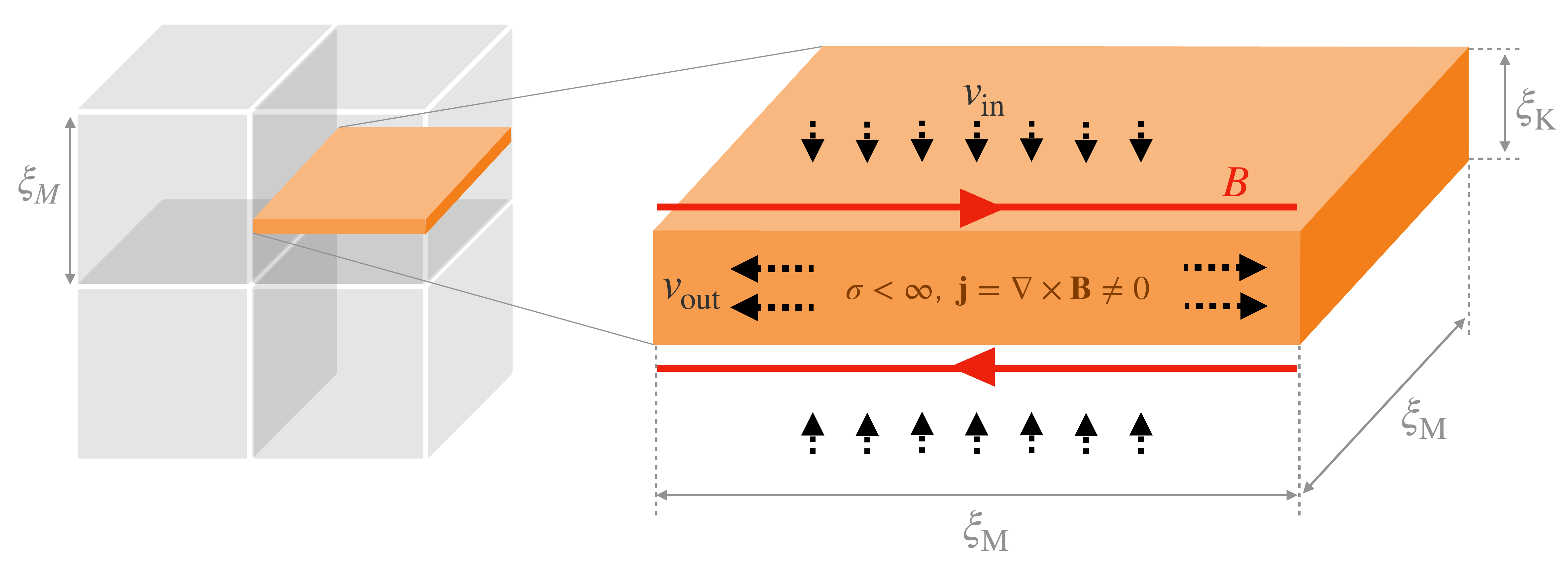}
\caption{\label{fig:SweetParker}An illustration of the current sheet (size of $\xi_{\rm  M}^{2}  \times  \xi_{\rm  K}$)
embedded in each patch (size of $\xi_{\rm  M}^3$). 
The magnetic reconnection occurs due to the finite electric conductivity on the current sheet.
The red solid and the black dashed arrows show the magnetic field and velocity flows, respectively. 
}
\end{figure*}

The Sweet--Parker reconnection mechanism is described as follows.
A current sheet of size $\xi_\text{M}^2\times\xi_\text{K}$ is formed at each boundary between two regions of coherent magnetic field lines, on which incoming magnetic field lines are dissipated, reconnect, and feed energy into the velocity field~\cite{sweet58, Parker57}.
See Fig.~\ref{fig:SweetParker}.
Let $v_\text{in}$ and $v_\text{out}$ denote incoming and outgoing velocities of material carrying the magnetic field lines, respectively, as illustrated in Fig.~\ref{fig:SweetParker}.
The mass conservation on each sheet implies $v_\text{in}\xi_\text{M}=v_\text{out}\xi_\text{K}$.
Comparing the inside and outside of the current sheet, the stationarity condition in the induction equation, Eq.~\eqref{eq:Faraday's induction equation}, is approximately $\frac{Bv_\text{in}}{\xi_\text{K}}\simeq \frac{B}{\sigma\xi_\text{K}^2}$, where we have approximated that ${\bm \nabla} \sim \xi_\mathrm{K}^{-1}$ at the current sheet. 
Therefore, we can express $v_\text{in}$ and $v_\text{out}$ in terms of $B, \xi_\text{M}$ and $\xi_\text{K}$ as
\begin{eqnarray}
    v_\text{in} = \frac{1}{\sigma\xi_\text{K}},\quad
    v_\text{out} = \frac{\xi_\text{M}}{\sigma\xi_\text{K}^2}.  \label{vinxik}
\end{eqnarray}
The outgoing flow spreads over the whole volume keeping the coherence length $\xi_\mathrm{K}$ while decreasing the amplitude.
Taking into account the dilution factor, we obtain
\begin{eqnarray}
    v^2
    \simeq \frac{\xi_\text{K}}{\xi_\text{M}}v_\text{out}^2.
    \label{eq:velocity_SP-reconnection}
\end{eqnarray}
Here we assume the high aspect ratio, $\xi_\text{M}/\xi_\text{K}\gg 1$, of the current sheet.
Since it takes the time,
\begin{eqnarray}
    \tau_\text{SP} := \frac{\xi_\text{M}}{v_\text{in}},
\end{eqnarray}
to process all the magnetic field within the volume $\xi_\text{M}^3$, the decay of the magnetic field energy in this regime proceeds, keeping the condition that the time scale equals to the conformal time at cosmic temperature $T$,
\begin{eqnarray}
    \tau(T) \left(= \tau_\text{SP}\right) = \sigma\xi_\text{M}\xi_\text{K}.
    \label{eq:Condition from the timescale of the Sweet--Parker}
\end{eqnarray}
Here we have used Eq.~\eqref{vinxik}.
This condition is the origin of the explicit time dependence of the evolution of the characteristic properties of the magnetic and velocity field. \\ 

Now let us take into account the effect of shear viscosity \cite{park+84} to determine the relation between the energy density of the magnetic and velocity fields.
Considering the energy budget along the fluid motion inside the current sheet, the energy of the magnetic field should be transferred to the energy of the outflow.
Taking into account the dissipation due to the shear viscosity, $\frac{\rho+p}{2}\frac{\eta v_\text{out}}{\xi_\text{K}^2}\xi_\text{M}$, 
which is evaluated by supposing the balance between the injection term and dissipation term in the Navier--Stokes equation, together with Eq.~\eqref{vinxik}, the energy conservation leads to another condition,
\begin{eqnarray}
    \frac{1}{2}B^2 = \frac{\rho+p}{2}\left(1+\text{Pr}_\text{M}\right)v_\text{out}^2
    \label{eq:Energy balance for Sweet--Parker with shear viscosity}
\end{eqnarray}
is imposed.
Here we have defined the magnetic Prandtl number,
\begin{eqnarray}
    \text{Pr}_\text{M}:=\sigma\eta. 
    \label{eq:Define the magnetic Prandtl number}
\end{eqnarray}
From Eqs.~\eqref{vinxik}, \eqref{eq:Condition from the timescale of the Sweet--Parker} and ~\eqref{eq:Energy balance for Sweet--Parker with shear viscosity}, we obtain a relation between $B$ and $\xi_{\rm M}$.
\begin{eqnarray}
    B^2 \tau^4 \simeq (\rho+p) \sigma^3\eta \xi_\text{M}^6, 
    \label{eq:constraint_nonlinear-shear_viscosity}
\end{eqnarray}
where we have used the approximation, $\text{Pr}_\text{M} \gg1$, which is the case in the early universe~\cite{Durrer+13}.

Combining Eqs.~\eqref{eq:Condition from conservation of the Hosking integral} and \eqref{eq:constraint_nonlinear-shear_viscosity}, we may determine the scaling behaviors of the magnetic field in this regime.
\begin{eqnarray}
    B &=& B_\text{ini}^{\frac{12}{17}} \,\xi_\text{M,ini}^{\frac{15}{17}} \left[(\rho+p) \sigma^3\eta\right]^{\frac{5}{34}} \tau^{-\frac{10}{17}},
    \label{eq:Magnetic field strength evolution for Sweet--Parker with shear viscosity}\\
    \xi_\text{M} &=& B_\text{ini}^{\frac{4}{17}} \,\xi_\text{M,ini}^{\frac{5}{17}} \left[(\rho+p) \sigma^3\eta\right]^{-\frac{2}{17}} \tau^{\frac{8}{17}}. 
    \label{eq:Magnetic field coherence length evolution for Sweet--Parker with shear viscosity}
\end{eqnarray}
Note that if the initial coherence length is large, 
\begin{equation}
    \tau < B_\mathrm{ini}^{-\frac{1}{2}} \xi_\mathrm{M,ini}^{\frac{3}{2}}  \left[(\rho+p) \sigma^3\eta\right]^{\frac{1}{4}},  \label{taufrozenA}
\end{equation}
the magnetic field is frozen and eventually starts the scaling evolution when the equality for Eq.~\eqref{taufrozenA} is satisfied.
As for the evolution of the velocity field, we find
\begin{eqnarray}
    v &=& \sigma^{\frac{1}{2}} \tau^{-\frac{3}{2}} \xi_\text{M}^2,\quad
    \xi_\text{K} = \sigma^{-1} \tau \xi_\text{M}^{-1},
    \label{eq:Velocity field coherence length evolution for Sweet--Parker with shear viscosity}
\end{eqnarray}
as the formulae describing the evolution in this regime.\\

The conditions for the system to be in this regime of the scaling evolution, Eqs.~\eqref{eq:Conditions of nonlinear regime with shear viscosity}, can be rewritten as
\begin{align}
    \text{Re}_\text{M} (\tau)
    &= \sigma^{\frac{5}{2}} \tau^{-\frac{5}{2}} \xi_\text{M}^5 (:=\text{Re}_\text{M}^{\text{SP}})
    \gg1,\; 
   \label{eq:The condition of low ReM for the Sweet--Parker with shear viscosity}    
    \\ 
    r_\text{diss}(\tau)
    &= \alpha \sigma^{-2} \eta^{-1} \tau^{2} \xi_\text{M}^{-2} (:=r_\text{diss}^{\text{SP}})
    \ll1,\quad\;\;\;
    \label{eq:The condition of low r_diss for the Sweet--Parker with shear viscosity}
\end{align}
which may be eventually violated so that the system enters another regime. Note that from these conditions, we can confirm the consistency of our solutions with the assumptions that the system is magnetically dominant, $B^2/[(\rho+p)v^2] = \text{Pr}_\text{M} \text{Re}_\text{M}^{\text{SP} \frac{2}{5}} \gg 1$, and that the aspect ratio is large, $\xi_\text{M}/\xi_\text{K} = \text{Re}_\text{M}^{\text{SP} \frac{2}{5}} \gg 
1$. 

\subsection{\label{sec:Nonlinear with drag force}Nonlinear regime with drag force}
For the case where the magnetic Reynolds number is larger than unity, while the drag force is dominant over the shear viscosity,
\begin{eqnarray}
    \text{Re}_\text{M}\gg1,\quad r_\text{diss}\gg1,
    \label{eq:Conditions of nonlinear regime with drag force}
\end{eqnarray}
the system is driven by the magnetic reconnection with dissipation due to the drag force.
To the best of our knowledge, this regime has never been discussed consistently with the other three regimes, so we work it out here.

Exactly the same discussion as the previous section, Sec.~\ref{sec:Nonlinear with shear viscosity}, holds (Eqs.~\eqref{vinxik} and \eqref{eq:Condition from the timescale of the Sweet--Parker}) up to the consideration about the energy balance.
On the other hand, the relation between the magnetic and kinetic energy (Eq.~\eqref{eq:Energy balance for Sweet--Parker with shear viscosity}) changes as follows.
In this regime, the dissipation of the kinetic energy  along the fluid motion inside the current sheet is replaced by the one due to the drag force, $\frac{\rho+p}{2}\alpha v_\text{out}\xi_\text{M}$.
Then, the energy balance leads to a condition, 
\begin{eqnarray}
    \frac{1}{2}B^2 \simeq \frac{\rho+p}{2}\alpha v_\text{out} \xi_\text{M}. 
    \label{eq:Energy balance for Sweet--Parker with drag force}
\end{eqnarray}
Here we have used the relation $\alpha \xi_\mathrm{M}/v_\mathrm{out} = r_\mathrm{diss} \mathrm{Pr_M} \gg 1$ so that the dissipation dominates over the remaining kinetic energy.
Using Eqs.~\eqref{vinxik} and \eqref{eq:Condition from the timescale of the Sweet--Parker}, Eq.~\eqref{eq:Energy balance for Sweet--Parker with drag force} can be rewritten as
\begin{eqnarray}
    B^2 \tau^2 = (\rho+p) \sigma \alpha \xi_\text{M}^4.
    \label{eq:constraint_nonlinear-drag_force}
\end{eqnarray}

Combining Eq.~\eqref{eq:Condition from conservation of the Hosking integral} and Eq.~\eqref{eq:constraint_nonlinear-drag_force}, we obtain
\begin{eqnarray}
    B &=& B_\text{ini}^{\frac{8}{13}} \,\xi_\text{M,ini}^{\frac{10}{13}} \left[(\rho+p) \sigma\alpha\right]^{\frac{5}{26}} \tau^{-\frac{5}{13}},
    \label{eq:Magnetic field strength evolution for Sweet--Parker with drag force}\\
    \xi_\text{M} &=& B_\text{ini}^{\frac{4}{13}} \,\xi_\text{M,ini}^{\frac{5}{13}} \left[(\rho+p) \sigma\alpha\right]^{-\frac{2}{13}} \tau^{\frac{4}{13}}. 
    \label{eq:Magnetic field coherence length evolution for Sweet--Parker with drag force}
\end{eqnarray}
Note that, once more, if the initial coherence length is large with
\begin{equation}
\tau < B_\mathrm{ini}^{-1} \xi_\mathrm{M,ini}^2  \left[(\rho+p) \sigma\alpha \right]^{\frac{1}{2}},  
\end{equation}
the magnetic field is frozen and starts the scaling evolution when the inequality is saturated.
The velocity field evolves in the same manner as described by Eqs.~\eqref{eq:Velocity field coherence length evolution for Sweet--Parker with shear viscosity} in this regime.\\

The Reynolds number and the ratio of dissipation terms are evaluated in the same way as Eqs.~\eqref{eq:The condition of low ReM for the Sweet--Parker with shear viscosity} and \eqref{eq:The condition of low r_diss for the Sweet--Parker with shear viscosity}.
The conditions for the system to be in this regime, Eqs.~\eqref{eq:Conditions of nonlinear regime with drag force}, can then be rewritten as
\begin{eqnarray}
    \text{Re}_\text{M}^{\text{SP}}
    \gg 1,\quad
    r_\text{diss}^{\text{SP}}
    \gg1.\quad\;\;
    \label{eq:The condition of low r_diss for the Sweet--Parker with drag force}
\end{eqnarray}

In this regime, we can also confirm the dominance of the magnetic energy as $B^2/[(\rho+p)v^2] = \text{Pr}_\text{M} \text{Re}_\text{M}^{\text{SP} \frac{2}{5}} r_\text{diss}^{\text{SP}} \gg 1$.
The large aspect ratio is confirmed in the same way as Sec.~\ref{sec:Nonlinear with shear viscosity}.

\subsection{\label{sec:Linear with shear viscosity}Linear regime with shear viscosity}
Next, we consider the case where the magnetic Reynolds number is smaller than unity and the shear viscosity is dominant over the drag force,
\begin{eqnarray}
    \text{Re}_\text{M}\ll1,\quad r_\text{diss}\ll1.
    \label{eq:Conditions of linear regime with shear viscosity}
\end{eqnarray}
In this case, the magnetic reconnection does not occur as the dominant contribution for the energy transfer between the magnetic and kinetic energy.
Instead, the velocity field is excited by the Lorentz force at the scale of the magnetic coherence length
\begin{eqnarray}
    \xi_\text{K}\simeq\xi_\text{M}.
    \label{eq:coherence lengths in the linear regimes}
\end{eqnarray}
The energy of the velocity field is brought to smaller scales by the Kolmogorov turbulence and finally dissipated by the shear viscosity.
The quasi-stationarity of the system implies the balance of the injection and the dissipation of the kinetic energy in the Navier--Stokes equation, Eq.~\eqref{eq:Navier--Stokes equation},
\begin{eqnarray}
    \frac{1}{\rho+p}\frac{B^2}{\xi_\text{M}}
    \simeq \eta \frac{v}{\xi_\text{K}^2},
\end{eqnarray}
from which one can express the typical velocity as
\begin{eqnarray}
    v = \frac{1}{\rho+p}\frac{B^2\xi_\text{M}}{\eta}.
    \label{eq:velocity in terms of magnetic field in the linear regime with shear viscosity}
\end{eqnarray}
For the Kolmogorov turbulence, it takes
\begin{eqnarray}
    \tau_\text{eddy}^\eta 
    := \frac{\xi_\text{K}}{v}
    = \frac{(\rho+p) \eta}{B^2}
\end{eqnarray}
to break the eddy of the coherence scale.
Therefore, the decay of the kinetic energy in this regime proceeds, keeping the condition
\begin{eqnarray}
    \tau(T) = \tau_\text{eddy}^\eta.
\end{eqnarray}

Combining Eqs.~\eqref{eq:Condition from conservation of the Hosking integral} and \eqref{eq:velocity in terms of magnetic field in the linear regime with shear viscosity}, we obtain the scaling behaviors of the magnetic field.
\begin{eqnarray}
    B &=& \left[(\rho+p)\eta\right]^{\frac{1}{2}}\tau^{-\frac{1}{2}},
    \label{eq:Magnetic field strength evolution for linear with shear viscosity}\\
    \xi_\text{M} &=& B_\text{ini}^{\frac{4}{5}} \,\xi_\text{M,ini}\left[(\rho+p)\eta\right]^{-\frac{2}{5}}\tau^{\frac{2}{5}}.
    \label{eq:Magnetic coherence length evolution for linear with shear viscosity}
\end{eqnarray}
The evolution of the velocity field are derived from Eqs.~\eqref{eq:coherence lengths in the linear regimes} and \eqref{eq:velocity in terms of magnetic field in the linear regime with shear viscosity}.
Note that if the initial magnetic field is too weak, 
\begin{equation}
\tau< \frac{(\rho+p)\eta}{B_\mathrm{ini}^2}, 
\end{equation}
the magnetic field is frozen and starts the scaling evolution when the inequality is saturated.
\\

The conditions for the system to be in this regime, Eqs.~\eqref{eq:Conditions of linear regime with shear viscosity}, can be rewritten as
\begin{align}
    \text{Re}_\text{M}(\tau)
    &= \frac{1}{\rho+p}\frac{\sigma B^2\xi_\text{M}^2}{\eta} (:= \text{Re}_\text{M}^\eta)\ll1,\\
    r_\text{diss}(\tau)
    &= \frac{\alpha\xi_\text{M}^2}{\eta} (:=  r_\text{diss}^\eta) \ll1,\quad
    \label{eq:The condition of linear and viscous}
\end{align}
from which the consistency of our solutions with the assumption that the system is magnetically dominant, $B^2/[(\rho+p)v^2] = \text{Pr}_\text{M} \left(\text{Re}_\text{M}^{\eta}\right)^{-1} \gg 1$, is confirmed.

\subsection{\label{sec:Linear with drag force}Linear regime with drag force}
Finally, we consider the case where the magnetic Reynolds number is smaller than unity and the drag force is dominant over the shear viscosity,
\begin{eqnarray}
    \text{Re}_\text{M}\ll1,\quad r_\text{diss}\gg1. 
    \label{eq:Conditions of linear regime with drag force}
\end{eqnarray}
In this case, the velocity field is excited by the Lorentz force and dissipated at smaller scales, similarly to the case in the previous section, Sec.~\ref{sec:Linear with shear viscosity}, and Eq.~\eqref{eq:coherence lengths in the linear regimes} also holds.
The quasi-stationarity of the system implies
\begin{eqnarray}
    \frac{1}{\rho+p}\frac{B^2}{\xi_\text{M}}\simeq \alpha v,
\end{eqnarray}
from which one can express the typical velocity, $v$, in terms of the typical strength, $B$, and the coherence length, $\xi_\text{M}$, of the magnetic field as
\begin{eqnarray}
    v = \frac{1}{\rho+p}\frac{B^2}{\alpha \xi_\text{M}}.
    \label{eq:velocity in terms of magnetic field in the linear regime with drag force}
\end{eqnarray}
It takes the time
\begin{eqnarray}
    \tau_\text{eddy}^\alpha 
    := \frac{\xi_\text{K}}{v}
    = \frac{(\rho+p) \alpha \xi_\text{M}^2}{B^2}
\end{eqnarray}
for the Kolmogorov turbulence to break the eddy of the coherence scale.
The decay of the kinetic energy in this regime proceeds, keeping the condition \cite{Banerjee+04},
\begin{eqnarray}
    \tau(T) = \tau_\text{eddy}^\alpha.
\end{eqnarray}

Combining Eqs.~\eqref{eq:Condition from conservation of the Hosking integral} and \eqref{eq:velocity in terms of magnetic field in the linear regime with drag force}, we obtain
\begin{eqnarray}
    B &=& B_\text{ini}^{\frac{4}{9}} \,\xi_\text{M,ini}^{\frac{5}{9}} \left[(\rho+p)\alpha\right]^{\frac{5}{18}} \tau^{-\frac{5}{18}},
    \label{eq:Magnetic field strength evolution for linear with drag force}\\
    \xi_\text{M} 
    &=& B_\text{ini}^{\frac{4}{9}} \,\xi_\text{M,ini}^{\frac{5}{9}} \left[(\rho+p)\alpha\right]^{-\frac{2}{9}} \tau^{\frac{2}{9}}.
    \label{eq:Magnetic field coherence length evolution for linear with drag force}
\end{eqnarray}
The evolution of the velocity field is derived from Eqs.~\eqref{eq:coherence lengths in the linear regimes} and \eqref{eq:velocity in terms of magnetic field in the linear regime with drag force}.
Note that if the initial coherence length is large, 
\begin{equation}
    \tau< \frac{(\rho+p) \alpha \xi_\mathrm{M,ini}^2}{B_\mathrm{M}^2}, 
\end{equation}
the magnetic field is frozen and starts the scaling evolution when the equality is satisfied. \\

The conditions for the system to be in this regime, Eqs.~\eqref{eq:Conditions of linear regime with drag force}, can be rewritten as
\begin{align}
    \text{Re}_\text{M} (\tau)
    &= \frac{1}{\rho+p}\frac{\sigma B^2}{\alpha} (:= \text{Re}_\text{M}^\alpha)\ll 1,\\
    r_\text{diss} (\tau) 
    &=\frac{\alpha\xi_\text{M}^2}{\eta} (:=r_\text{diss}^\alpha) \gg1,\quad
    \label{eq:The condition of linear and dragged}
\end{align}
from which the consistency of our solutions with the assumption that the system is magnetically dominant, $B^2/[(\rho+p)v^2] = \text{Pr}_\text{M} (\text{Re}_\text{M}^{\alpha})^{ -1} r_\text{diss}^\alpha \gg 1$, is confirmed.\\

Note that the linear regimes with both shear viscosity (Sec.~\ref{sec:Linear with shear viscosity}) and drag force (Sec~\ref{sec:Linear with drag force}) were studied in Ref.~\cite{Banerjee+04}, and it was claimed that the system in the linear regimes with shear viscosity is frozen in the realistic situation.
On the contrary, here we have performed a general study, not restricting ourselves to the situation which was discussed in the literature.
In Sec.~\ref{sec:Discussion}, we will argue that 
the system can evolve according to the scaling law we derived 
also in realistic situation.

\section{\label{sec:Integartion of the analysis}Integration of the analyses}
In the previous section, we have conducted regime-dependent analyses.
Let us confirm the consistency, by showing that the solutions coincide at the boundary of linear and non-linear regimes.

First, we focus on the properties of the magnetic and velocity field in the scaling evolution regimes. On the coherence length, we have
\begin{eqnarray}
    \xi_\text{M}=\xi_\text{K},\quad \text{when}\quad \text{Re}_\text{M}=1
    \label{eq:Relation between magnetic and kinetic coherence length}
    ~~~~
\end{eqnarray}
in all regimes. 
We identify $v$ as the representative velocity in each regime. 
For the non-linear regime, we use characteristic velocity in the Sweet--Parker reconnection. We may confirm that these expressions match velocities for linear regimes on the boundary between the linear and non-linear regimes,  
\begin{eqnarray}
  v =  
  \left\{
    \begin{matrix}
    \frac{1}{\rho  +  p}  \frac{B^2  \xi_\text{M}}{\eta},
    &
    \text{($\text{Re}_\text{M}=1$, viscous regimes)}
    ~
    \\
    \frac{1}{\rho  +  p}  \frac{B^2}{\alpha  \xi_\text{M}},
    ~~
    &
    \text{($\text{Re}_\text{M}=1$, dragged regimes)}
    \end{matrix}
    \right.
    \label{eq:Relations between velocity}
\end{eqnarray}
Using these expressions of $v$ for each regime, we may express
\begin{eqnarray}
    \frac{B^2}{\rho+p}=
    \left\{\begin{matrix}
    \text{Pr}_\text{M} v^2,
    ~~~~~~~
    \text{($\text{Re}_\text{M}=1$, viscous regimes)}
    ~\,
    \\
    \text{Pr}_\text{M}r_\text{diss} v^2.
    ~~
    \text{($\text{Re}_\text{M}=1$, dragged regimes)}
    \end{matrix}
    \right.\quad
    \label{eq:Relations between magnetic and kinetic energy density}
\end{eqnarray}

Also, we confirm that the time scales of the evolution of the system are connected smoothly at the boundary between the linear and non-linear regimes,
\begin{eqnarray}
    \tau_\text{SP} = 
    \left\{\begin{matrix}
    \tau_\text{eddy}^\eta,
    &
    \text{($\text{Re}_\text{M}=1$, viscous regimes)}
    ~
    \\
    \tau_\text{eddy}^\alpha, 
    &
    \text{($\text{Re}_\text{M}=1$, dragged regimes)}
    \end{matrix}\right.\quad
    \label{eq:Relations between time scales}
\end{eqnarray}
which remains identical to the conformal time of the universe, $\tau(T)$, in the scaling regime.
This also suggests that the condition for the system to be frozen is also connected smoothly. 
Note that these expressions, Eqs.~\eqref{eq:Relations between velocity}-\eqref{eq:Relations between time scales}, are clearly self-consistent at the boundary of the viscous and dragged regimes ($r_\text{diss}=1$).
We conclude that the decay laws in linear and non-linear regimes 
as well as viscous and dragged regimes are consistent at their boundary ($\text{Re}_\text{M} = 1$ and $r_\text{diss}=1$).\\

Now let us summarize the decay laws in Table \ref{tab:summary of the decay laws}.
Detailed discussions are found in the sections shown in the first column.
The state of the system is classified into four regimes, according to the criteria written in the second-to-fourth columns. In each regime, the formulae describing the decay laws are the equations specified in the fifth and sixth columns. The decay time scales are shown in the last column. When the conformal time of the universe $\tau(T)$ is smaller than these time scales, the decay processes are too slow to operate, and the system is frozen until when $\tau(T)$ catches up with the decay time scales. 
We have confirmed that each regime including the condition to be frozen is connected smoothly. 
Therefore, we may continuously evolve the system, even involving multiple regimes, by following the decay laws in  Table~\ref{tab:summary of the decay laws}.

\begin{table*}
\caption{\label{tab:summary of the decay laws}Summary of the decay laws of non-helical and magnetically dominant regimes.}
\begin{ruledtabular}
\begin{tabular}{c|ccc|ccc|c}
 & Regimes & Dissipation & Condition & \multicolumn{2}{c}{Decay laws of the magnetic field} & velocity field & Decay time scale\\ \hline
 &&&&&&&\vspace{-3mm}\\
 Sec.~\ref{sec:Nonlinear with shear viscosity} &\multirow{2}{*}{Nonlinear} & Shear viscosity & Eqs.~\eqref{eq:The condition of low r_diss for the Sweet--Parker with shear viscosity} & Eqs.~\eqref{eq:Magnetic field strength evolution for Sweet--Parker with shear viscosity} and \eqref{eq:Magnetic field coherence length evolution for Sweet--Parker with shear viscosity} & or frozen & \multirow{2}{*}{Eqs.~\eqref{eq:Velocity field coherence length evolution for Sweet--Parker with shear viscosity}} & \multirow{2}{*}{$\tau_\text{SP} = \sigma\xi_\text{M}\xi_\text{K}$}\\
 Sec.~\ref{sec:Nonlinear with drag force} && Drag force & Eqs.~\eqref{eq:The condition of low r_diss for the Sweet--Parker with drag force} & Eqs.~\eqref{eq:Magnetic field strength evolution for Sweet--Parker with drag force} and \eqref{eq:Magnetic field coherence length evolution for Sweet--Parker with drag force} & or frozen &&\\
 Sec.~\ref{sec:Linear with shear viscosity} &\multirow{2}{*}{Linear} & Shear viscosity & Eqs.~\eqref{eq:The condition of linear and viscous} & Eqs.~\eqref{eq:Magnetic field strength evolution for linear with shear viscosity} and \eqref{eq:Magnetic coherence length evolution for linear with shear viscosity} & or frozen & Eqs.~\eqref{eq:coherence lengths in the linear regimes} and \eqref{eq:velocity in terms of magnetic field in the linear regime with shear viscosity} & \multirow{2}{*}{$\left.\begin{matrix}\tau_\text{eddy}^\eta \\ \tau_\text{eddy}^\alpha\end{matrix}\right\} =\dfrac{\xi_\text{K}}{v}$}\\
 Sec.~\ref{sec:Linear with drag force} && Drag force & Eqs.~\eqref{eq:The condition of linear and dragged} & Eqs.~\eqref{eq:Magnetic field strength evolution for linear with drag force} and \eqref{eq:Magnetic field coherence length evolution for linear with drag force} & or frozen & Eqs.~\eqref{eq:coherence lengths in the linear regimes} and \eqref{eq:velocity in terms of magnetic field in the linear regime with drag force} &
\end{tabular}
\end{ruledtabular}
\end{table*}

\section{\label{sec:Discussion}Discussion}
In this letter, we have provided a comprehensive analysis that describes the evolution of the magnetically dominant and non-helical magneto-hydrodynamic system.
We limit ourselves to the cases when the magnetic field is not very strong so that the Lundquist number is not very large and that the Sweet--Parker reconnection (but not the fast reconnection) is relevant.
The results are summarized in Table~\ref{tab:summary of the decay laws}.
Importantly, our analysis predicts quite different evolution history of the primordial magnetic field, compared with the one by Banerjee and Jedamzik \cite{Banerjee+04}.
The difference mainly comes from the fact that the approximately conserved Hosking integral was not taken into account there, which was first pointed out by Hosking and Schekochihin \cite{Hosking+21}.

\begin{figure}[b]
\includegraphics[width=0.45\textwidth]{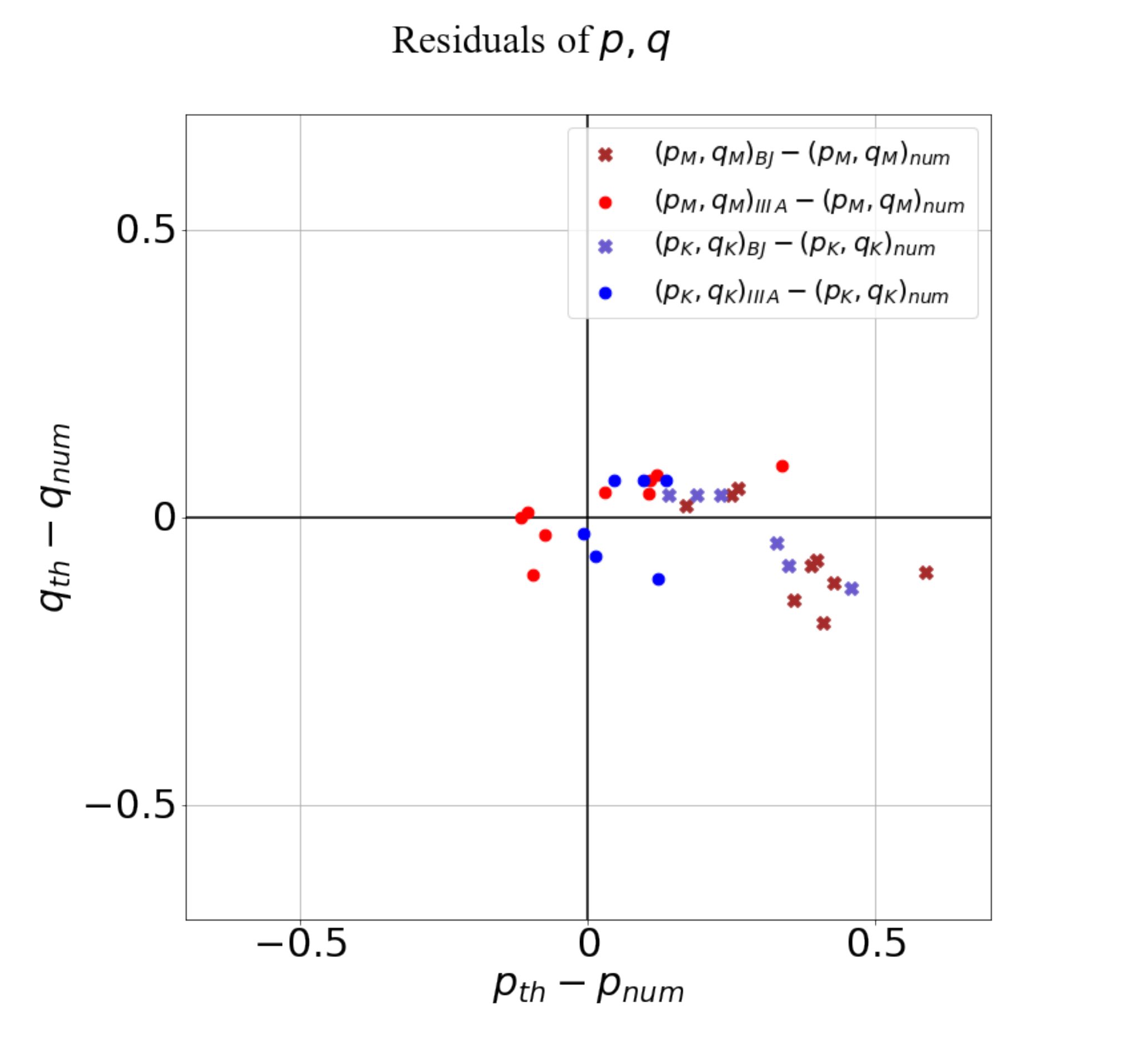}
\caption{\label{fig:Residuals} Comparison between theoretical analyses (dots: the analysis in Sec.~\ref{sec:Nonlinear with shear viscosity}; crosses: Banerjee and Jedamzik \cite{Banerjee+04}) and the numerical results in the literature \cite{Brandenburg+17,zhou2022scaling, Brandenburg:2016odr}. 
Since the power indices read off from numerical simulations may well contain errors around $\pm 0.1$, one should conclude that the theoretical model reproduces the numerical simulation well if the points are located
within a circle of a radius $\sim 0.14$ centered at the origin.}
\end{figure}

Let us see how the analysis with the Hosking integral better describes the results of existing numerical MHD simulations compared with the one by Banerjee and Jedamzik \cite{Banerjee+04}.
See the plot in Fig.~\ref{fig:Residuals}.
By parametrizing time dependence of each quantity as~\cite{Brandenburg:2016odr}
\begin{eqnarray}
    \frac{B^2}{\rho+p}\propto \tau^{-p_\text{M}},\;
    \xi_\text{M}\propto \tau^{q_\text{M}},\;
    v^2\propto \tau^{-p_\text{K}},\;
    \xi_\text{K}\propto \tau^{q_\text{K}},\quad
\end{eqnarray}
we compare the analytic formula,
which is first studied in Ref.~\cite{Hosking+21} and summarized in Sec.~\ref{sec:Nonlinear with shear viscosity} of the present letter, and the numerical results in the literature \cite{Brandenburg+17, zhou2022scaling, Brandenburg:2016odr}.
The five runs that can be interpreted to be magnetically dominant and non-helical from Ref.~\cite{Brandenburg+17}, the non-helical one in Ref.~\cite{Brandenburg:2016odr}, and the three with the standard magnetic dissipation term in Ref.~\cite{zhou2022scaling} are employed.
Plots near the origin indicate that the theory well describes the numerical calculations.
The dots representing our analysis scatter around the origin\footnote{The apparent outlier (the red point in the above-right of the origin) is the run ``K60D1c'' in Zhou et al \cite{zhou2022scaling}. We do not have a clear explanation about the origin of the deviation, but could be originated from the insufficient resolution.}, while the crosses representing the analysis in Ref.~\cite{Banerjee+04} are off the origin rightward.
Note that we take values of the parameters, $p_\mathrm{M/K}$ and $q_\mathrm{M/K}$, up to their fluctuations in time (which may be roughly $\lesssim 0.1$), from the results of numerical simulations in the literature, and they are not so accurate.
Nevertheless, this plot strongly suggests that the new analysis with the Hosking integral better describes the reality, compared with the one in Banerjee and Jedamzik \cite{Banerjee+04}, which tends to predict too fast decay of both the magnetic and the kinetic energy.

\begin{figure}[h]
\includegraphics[width=0.475\textwidth]{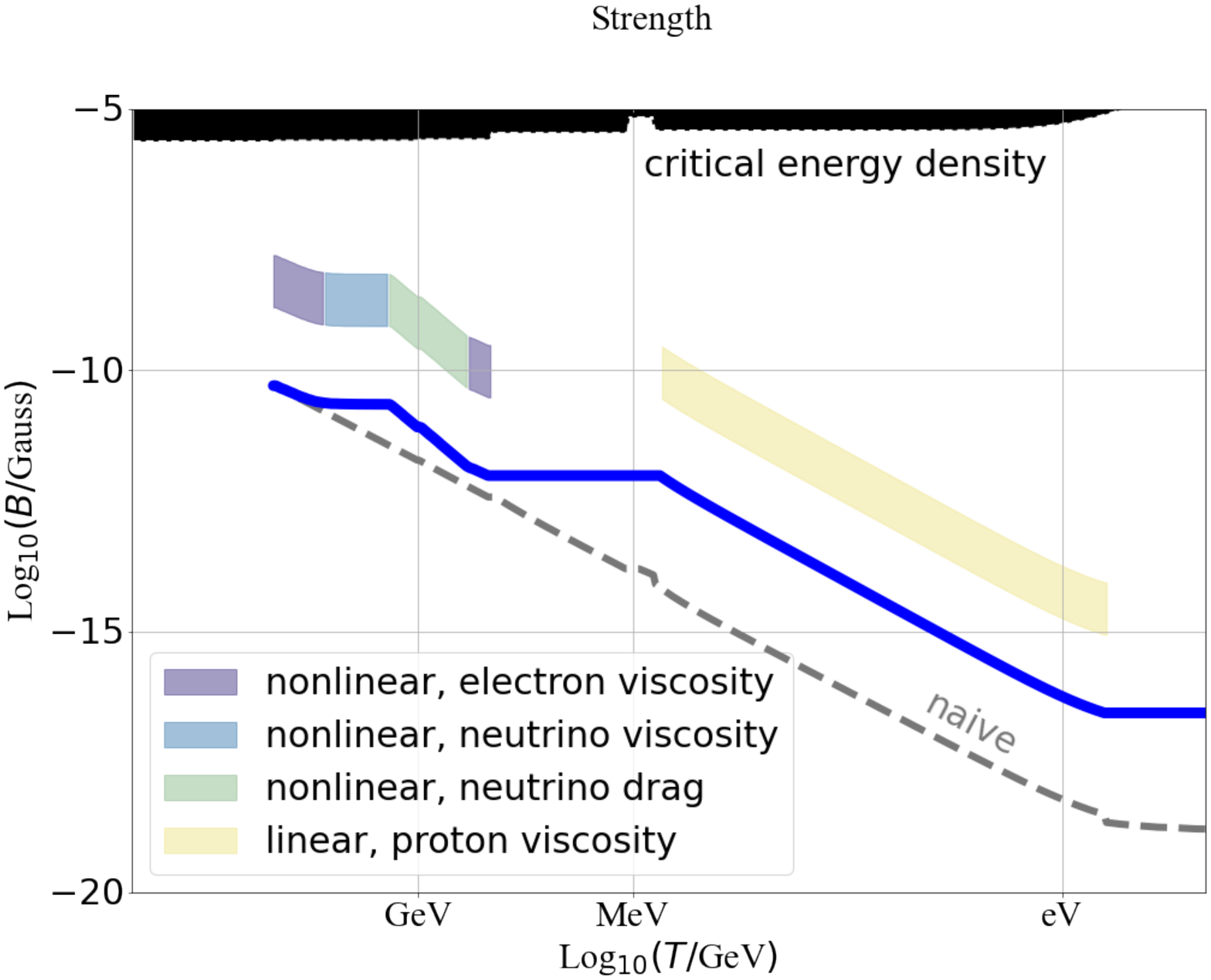}
\caption{\label{fig:Strength} The evolution of the strength of the magnetic field. Solid lines are based on our analysis, while the dashed line is based on the naive extrapolation of the analysis by Banerjee and Jedamzik \cite{Banerjee+04}. Colored bars indicate which regime determines the scaling evolution at each temperature range. The black shaded region corresponds to over-closure of the universe.}
\end{figure}
\begin{figure}[h]
\includegraphics[width=0.475\textwidth]{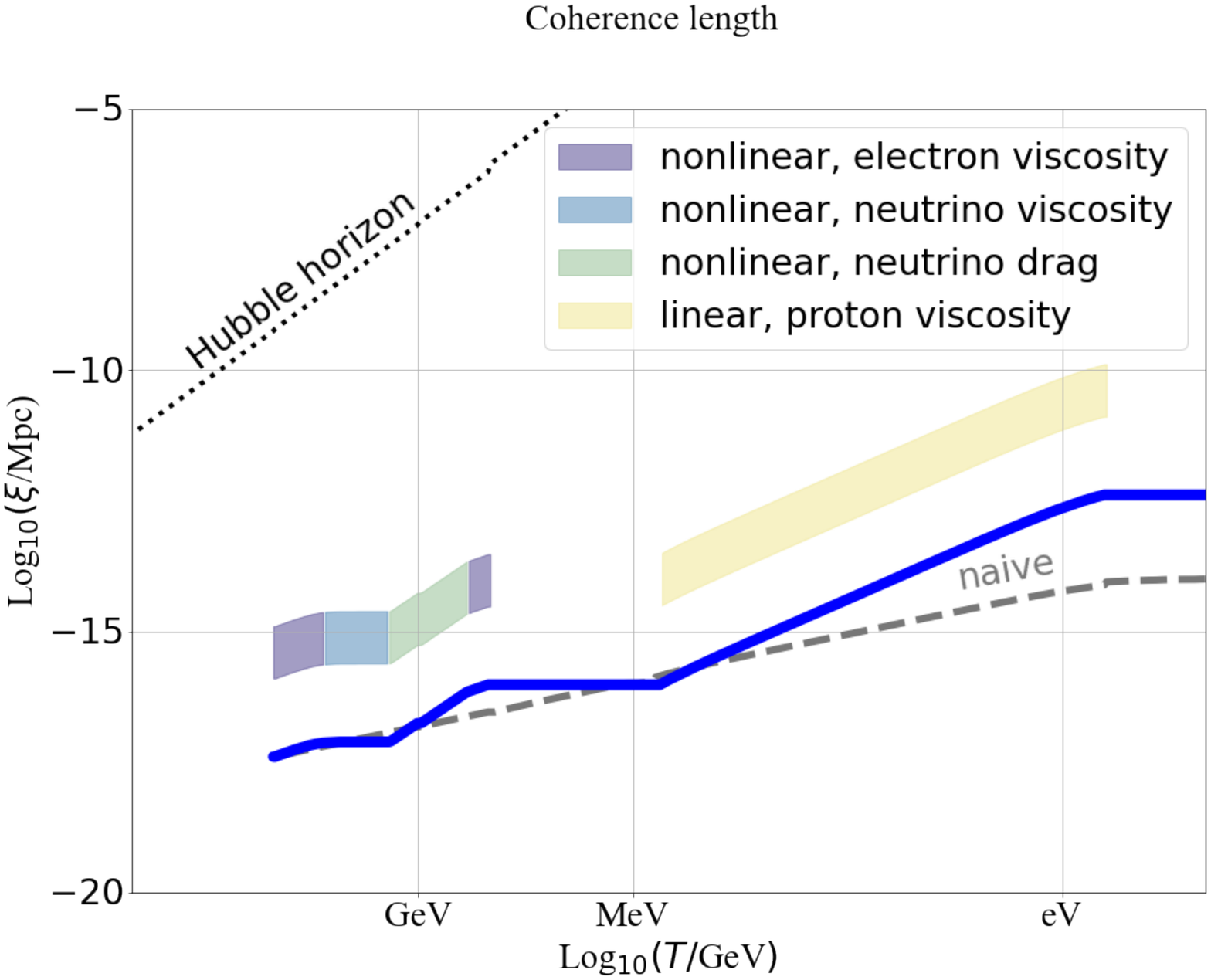}
\caption{\label{fig:Length} The evolution of the coherence length of the magnetic field. Solid lines are based on our analysis, while the dashed line is based on the naive extrapolation of the analysis by Banerjee and Jedamzik \cite{Banerjee+04}. Colored bars indicate which regime determines the scaling evolution at each temperature range. The dotted line indicates the Hubble horizon.}
\end{figure}

Now we turn to the implication of the main topic of this letter, the cosmological evolution of the magnetic field that involves not only the non-linear regime but also the linear regime with dissipation due to the shear viscosity as well as dissipation due to the drag force.
As a demonstration, we investigate the cosmological evolution of the magnetic field generated at the time of electroweak symmetry breaking.
In this case, we need to take into account the shear viscosity due to electron-electron, electron-neutrino, proton-proton, electron-photon, and hydrogen-hydrogen collisions and the drag force due to the free-streaming neutrinos and photons, when their mean free path is larger than the relevant length scale of the system \cite{arnold2000transport, Banerjee+04}.
The Hubble friction in the matter domination and the ambipolar drag \cite{Banerjee+04} also serve as the drag forces.
These phenomena lead to the change of the regimes, where we can smoothly connect the evolution of magnetic field by using Table~\ref{tab:summary of the decay laws}.

Figures \ref{fig:Strength} and \ref{fig:Length} show an example of the evolution history of magnetic field strength and coherence length, respectively.
The initial condition is set such that the energy density of the magnetic field is several orders below the critical energy density, $B_\text{ini} = 10^{-10.3}\;\text{G}$, and that the magnetic coherence length is safely short, $\xi_\text{M,ini}=10^{-17.4}\;\text{Mpc}$, at the electroweak symmetry breaking temperature $\sim 100\;\text{GeV}$.
Note that if we assume that the primordial magnetic field is generated before the electroweak symmetry breaking, such a choice of the initial condition is motivated by the constraint that comes from the big-bang nucleosynthesis \cite{kamada+21}.
A stronger or longer-ranged $\text{U}(1)_Y$ magnetic field unavoidably generates baryon isocurvature perturbation through the chiral anomaly at the electroweak symmetry breaking, leading to the deuterium overproduction at the big-bang nucleosynthesis to be inconsistent with the present universe.
The initial condition chosen here almost saturates this constraint.

With this initial condition of the magnetic field, the reconnection is driven by the Sweet--Parker mechanism at the non-linear stage, and thus the analysis in the present letter applies.
At first, the system is in the nonlinear regimes for a while.
Since the dimensionful Fermi constant $G_\text{F}$ is involved in the electron-neutrino collision, the dominant contributor to the shear viscosity eventually changes from electron-electron collisions
to electron-neutrino collisions.
Since the neutrino viscosity evolves as $\eta_{\nu}  \propto  \tau^4$, which compensates the $\tau$ dependence in Eqs.~\eqref{eq:Magnetic field strength evolution for Sweet--Parker with shear viscosity} and \eqref{eq:Magnetic field coherence length evolution for Sweet--Parker with shear viscosity}, the time evolution is accidentally frozen.
When the neutrinos start free-streaming, neutrino drag begins to dissipate the energy of the system.
Noting that the neutrino drag $\alpha_\nu  \propto  \tau^{-4}$, the magnetic coherence length increases following Eq.~\eqref{eq:Magnetic field coherence length evolution for Sweet--Parker with drag force}. 
Therefore, $r_\text{diss}$ decreases and the system enters the nonlinear regime with the electron viscosity again.
At around the QCD phase transition, the system becomes frozen due to the sudden change of the viscosity, which becomes dominated by the proton-proton collision.
The scaling evolution resumes when electrons become massive and the photon drag coefficient decreases.
Then $r_\text{diss}$ drops below unity, and the linear regime with proton viscosity begins.
After the recombination, the system becomes frozen again.

Let us demonstrate how our analysis is different from what is commonly accepted.
According to the analysis by Banerjee and Jedamzik \cite{Banerjee+04}, the evolution in the turbulent regime is $B\propto \tau^{-5/7}$ and $\xi_\text{M}\propto \tau^{2/7}$, assuming the Batchelor spectrum.
This scaling behavior is often just extrapolated throughout the history before the recombination, to connect initial conditions and resultant configurations.
We plot these naive extrapolations with dashed lines in Figs.~\ref{fig:Strength} and \ref{fig:Length}.

One can see that the evolution history is quite different.
First, based on our analysis, the resultant strength is much stronger and the coherence length is much longer than the previous expectations. 
This is because the conserved quantity, the Hosking integral, leads to a slower decay of the magnetic energy with the inverse cascade.
Note that the resultant properties of magnetic field are not likely to account for the long-range intergalactic magnetic field suggested by the blazar observations~\cite{2020ApJ...902L..11A} although the magnetic field strength becomes stronger than the previous estimate.

Second, it is impossible to approximate the evolution throughout the history before the recombination by a single power law.
This is because the time-dependent dissipation coefficients, $\sigma, \eta,$ and $\alpha$, play critical roles even in the nonlinear regime.

It is essential that the magnetic field is often frozen in its evolution history. When it is frozen, the initial conditions and the decay time scale at that time are insufficient to determine the properties at that time. 
To connect arbitrary initial conditions and the corresponding final configurations, one should draw the evolution history like Figs.~\ref{fig:Strength} and \ref{fig:Length} for every initial condition, since the evolution histories are highly diverse depending on the initial conditions. 
Since the system experiences multiple regimes in general, integration of regime-dependent analysis given in Sec.~\ref{sec:Integartion of the analysis} is inevitable to follow this evolution.

In summary, we have clarified decisive elements to determine the evolution history of the cosmological magnetic field in the case it is non-helical and its energy surpasses the kinetic energy.  As a result, the evolution may be classified into the four possible regimes summarized in Table \ref{tab:summary of the decay laws}.

Cases with helical magnetic field and kinetically dominant regimes require some extension of the analysis.
Also, we have only considered the magnetic reconnection described by the Sweet--Parker model \cite{sweet58, Parker57}.
Different mechanisms of magnetic reconnection \cite{ji2011phase} should also be taken into account for full description \cite{Hosking+22}.
Our forthcoming paper \cite{FKUYinPrep} will give a more general description about the evolution of the cosmological magneto-hydrodynamic system.

\begin{acknowledgments}
    The work of FU was supported by the Forefront Physics and Mathematics Program to Drive Transformation (FoPM).
        The work of MF was supported by the Collaborative Research Center SFB1258 and by the Deutsche Forschungsgemeinschaft (DFG, German Research Foundation) under Germany's Excellence Strategy - EXC-2094 - 390783311, JSPS Core-to-Core Program (No.JPJSCCA20200002).
This work was partially supported by JSPS KAKENHI, Grant-in-Aid for Scientific Research Nos.\ (C)JP19K03842(KK), (S)20H05639(JY), the Grant-in-Aid for Innovative Areas Nos.\ 18H05542(MF) and 20H05248(JY). 
\end{acknowledgments}

\bibliographystyle{unsrt}
\bibliography{Fujiwara_Kamada_Uchida_Yokoyama}
\end{document}